\documentstyle[12pt]{article}
\def\be{\begin{equation}}
\def\ee{\end{equation}}

\def\sp{\hskip2pt}
\def\li{q\to-1}
\def\ff{f(\theta)}
\def\drr{{d\over d\theta}}

\def\bb #1 {\vskip0pt\noindent[#1]\hskip5pt}
\def\r #1 {\hskip5pt[#1]\hskip3pt}

\def\pt{{\partial\over\partial\theta}}
\def\dt{{\partial\over \partial t}}

\def\dz{\partial_t}

\def\qm{\lim_{q\to-1}}

\def\ff{f(\theta)}

\def\ss{{\sum_{m=0}^\infty}}
\def\d{{{\cal D}_L}}
\def\dR{{{\cal D}_R}}
\def\ie{{\it i.e.}}
\catcode`@=11
\newdimen\z@ \z@=0pt
\def\m@th{\mathsurround=\z@}
\def\ialign{\everycr{}\tabskip\z@skip\halign} 
\def\eqalign#1{\null\,\vcenter{\openup\jot\m@th
  \ialign{\strut\hfil$\displaystyle{##}$&$\displaystyle{{}##}$\hfil
      \crcr#1\crcr}}\,}
\def\matrix#1{\null\,\vcenter{\normalbaselines\m@th
    \ialign{\hfil$##$\hfil&&\quad\hfil$##$\hfil\crcr
      \mathstrut\crcr\noalign{\kern-\baselineskip}
      #1\crcr\mathstrut\crcr\noalign{\kern-\baselineskip}}}\,}

\@addtoreset{equation}{section}

\catcode`@=12

\begin{document}
\begin{flushright}
DAMTP/96-52
\\
FTUV 96-27 / IFIC 96-31
\end{flushright}
\vspace{1cm}
\begin{center}
\begin{Large}
{\bf Supersymmetry from a braided point of view}
\end{Large}
\\
\vspace{1cm}
{\bf R.S. Dunne$^\dagger$, A.J. Macfarlane$^\dagger$, 
J.A. de Azc\'arraga$^\ddagger$
and J.C. P\'erez Bueno$^\ddagger$
\footnote{e-mails: r.s.dunne@damtp.cam.ac.uk; a.j.macfarlane@damtp.cam.ac.uk;
\\ 
$\phantom{aaaaaaaaa:: }$azcarrag@evalvx.ific.uv.es; pbueno@lie.ific.uv.es}}
\vskip10pt
{\it $^\dagger$ Department of Applied Mathematics \& Theoretical Physics}\vskip0pt
{\it University of Cambridge, Cambridge CB3 9EW}
\\
\vskip10pt
{\it $^\ddagger$ Departamento de F\'{\i}sica Te\'orica and IFIC,}\vskip0pt
{\it Centro Mixto Universidad de Valencia-CSIC}\vskip0pt
{\it E-46100-Burjassot (Valencia) Spain.}
\end{center}
\abstract
{\noindent
We show that one-dimensional superspace is isomorphic to a non-trivial but
consistent limit as 
$q\to-1$ of the braided line. 
Supersymmetry is identified as translational invariance along this line. 
The supertranslation generator and covariant derivative
are obtained in the limit in question as the
left and right derivatives of the 
calculus on the braided line.}
\vskip0pt
\medskip

\section{Introduction}

Supersymmetry \cite{Fer}
has attracted a great deal of interest from physicists and mathematicians. 
It finds numerous applications in particle physics and statistical mechanics, 
in particular in the
theories of supersymmetric quantum mechanics 
\cite{Witten,CR}, superstrings and supergravity \cite{Fer,GSW}. 
There also exist `super' 
extensions to many of the structures and concepts from ordinary mathematics. 
A particularly useful concept in dealing with supersymmetry is the idea of 
(rigid) 
superspace \cite{SS,FWZ} which is usually introduced as the coset
[super-Poincar\'e/Lorentz], and which may be parametrized by the ordinary 
spacetime coordinates $\{x^\mu\}$ supplemented by 
Grassmann variables $\{\theta_\alpha\}$ transforming as a spinor 
(Majorana for example).
We consider here only the simplest case of one-dimensional superspace.

Of central importance in supersymmetric theories is the $Z_2$-graded 
structure which permeates them. Their associated superalgebras are most
naturally represented in superspace, using generators built out of the 
superspace coordinates and the corresponding derivatives.   
Supersymmetry transformations which mix the odd and even sectors of superspace 
are generated by the odd elements of the super-Poincar\'e algebra. 
In one-dimensional supersymmetry, or `supermechanics' 
\cite{Witten,CR}, the superalgebra
is just $\{Q,Q\}=2Q^2=2H\,,\,[Q,H]=0$, 
where $H$ is the Hamiltonian and $Q$ the supercharge. 
In this case superspace
consists of an even (time) coordinate $t$ and a single 
real Grassmann variable 
$\theta$, satisfying $[t,\theta]=0$ and $\theta^2=0$.  $\delta t H$ generates a
time translation, while $\epsilon Q$ generates a supertranslation $\epsilon$ 
and also produces a time shift $i\epsilon\theta$, so that under a general 
transformation
\be
\theta\mapsto\theta+\epsilon\quad,
\label{onea}
\ee
\be
t\mapsto t+\delta t+i\epsilon\theta\quad,
\label{oneb}
\ee
where $\epsilon^2=0$ and $\{\epsilon,\theta\}=\epsilon\theta+\theta\epsilon=0$.

The aim of this paper is to provide a novel geometric interpretation of 
supersymmetry and superspace, 
showing how they arise in the $q\to-1$ limit of the 
simplest braided Hopf algebra \cite{MajI,MajII}, the braided line, 
which we review in section 3 along with its associated calculus. 
The braided line is characterized by a single deformation parameter $q$. 
Its algebra contains a single variable
$\theta$, and (for $q$ not a root of unity) there are no additional relations, 
but it has nontrivial braiding in its
tensor product structure. 
It is well known \cite{MajIII} that when we set $q=-1$, the braided line 
reduces to the superline, consisting of a single Grassmann variable, \ie\ we 
then have the relation $\theta^2=0$ as above. 
From this point of view the transformation (\ref{onea}) 
is just a translation along the superline. 
We show in section 4 that, if instead of setting $q=-1$, we work initially 
with generic $q$ ($q$ not equal to a root of unity), and then pass to the limit
as $q\to-1$ in a non-trivial but fully consistent way,
some extra structure emerges. 
In fact our work leads us to the following conclusion: 
the one-dimensional superspace $(t,\theta)$ coincides in the
limit in question with the braided line, the supersymmetry transformation 
generated by $Q$  corresponds to a 
translation along this line, and a theory is supersymmetric if it is invariant 
under such translations.

Of particular interest is the fact that in  this limit the braided line algebra,
 suitably normalized, can be separated 
into two parts, one described by a Grassmann variable $\theta$, the other 
by an ordinary even variable $t$. 
This separation cannot be extended to the rest of the Hopf algebraic 
structure associated with
the braided line, and this is reflected in the fact that under a
translation along this line $\theta$ and $t$ transform as in (\ref{onea})
and (\ref{oneb}). 
Since our $q\to-1$ limit of the braided line cannot be written as 
into the tensor product of a $t$ part and a $\theta$ part, the
proper way to view superspace is as a single geometric entity, 
not as a composite.
  
When we take the limit of calculus on the braided line as $q\to-1$, 
we are able to
identify the left derivative with the supercharge $Q$, and the right 
derivative with the covariant derivative $D$ (see also \cite{AldAz,AM}), 
obtaining the familiar superspace realizations of these operators by means 
of a chain rule expansion.

In fact our work here is the simplest case of a much more general result. 
In a forthcoming paper \cite{DMPA} 
we extend our work to the case of $q$ an arbitrary root of unity, obtaining
numerous results of interest in fractional supersymmetry, 
as well as developing a complete theory of $q$-integration in this limit. This $q^n=1$ case
involves a $Z_n$
grading, which necessitates the introduction of an ordering amongst the various elements, making it
qualitatively more complicated than the $Z_2$ case treated here. 
We are also preparing a paper \cite{Dun} in which similar techniques are used to show how
internal spin arises naturally in certain limits of the $q$-deformed angular momentum group.

\section{Brackets and q-grading}

In this section we establish the bracket notation and grading scheme 
to be used throughout;
$q$ and $r$ are just arbitrary complex numbers. 
We begin by defining the bracket
\be
[A,B]_{q^r}:=AB-q^r BA
\quad.\label{two}
\ee
If we assign an integer grading $g(X)$ to each element $X$ of some algebra, 
such that $g(1)$=0 and
$ g(XY)=g(X)+g(Y),$ 
for any $X$ and $Y$, we can define a bilinear graded $z$-bracket as follows,
\be 
[A,B]_z=AB-q^{-g(A)g(B)}BA\quad,\quad z=q^{-g(A)g(B)}
\quad.
\label{four}
\ee
Here $A$ and $B$ are elements of the algebra, and hence of pure grade. 
The definition can be extended to mixed grade terms using 
bilinearity. 
We will also make use of the following notation,
\be 
[r]_q={{1-q^r}\over{1-q}}\quad,\quad
[r]_q!=[r]_q[r-1]_q[r-2]_q...[2]_q[1]_q\quad,
\label{six}
\ee
supplemented by $[0]_q!=1$. 
Note that when $q=-1$ our grading scheme becomes degenerate, so that in 
effect the grading of an element is only defined modulo $2$. 
If $q=-1$ we also have $[r]_q=0$ when $r$ modulo $2$ is zero $(r\neq0)$. 
A similar situation
arises whenever $q$ is 
an arbitrary root of unity, but in this paper we 
restrict our attention to $q=-1$.

\section{$q$-calculus and the braided line}
The simplest example of a braided Hopf algebra \cite{MajI,MajII} 
is the braided line, a deformation of the ordinary line
characterized by a single parameter $q$.
We consider a 
braided line algebra consisting of a single variable $\theta$, of grade 1, 
upon which no additional conditions are placed
for generic $q$ (by which we mean $q$ not a root of unity). 
The Hopf structure of this deformed line is as follows.
It has braided coproduct,
\be
\Delta\theta=\theta\otimes1+1\otimes\theta\quad,
\label{seven}
\ee
where the braiding is given by
\be
(\theta^a\otimes \theta^b)(\theta^c\otimes \theta^d)=
q^{bc}\theta^{a+c}\otimes \theta^{b+d}\quad,
\label{eight}
\ee 
so that 
\be
\eqalign{
(1\otimes\theta)(\theta\otimes1)&=q\theta\otimes\theta\quad,
\quad
(\theta\otimes1)(1\otimes\theta)=\theta\otimes\theta\quad.\cr}
\label{nine}
\ee
There are also a counit and antipode,
\be
\eqalign{\varepsilon(\theta)&=0\quad,\quad
S(\theta^r)=q^{r(r-1)\over2}(-\theta)^r\quad,\cr}
\label{oneone}
\ee
which satisfy all the usual Hopf algebraic relations,
as long as the braiding is remembered. 
From the braided Hopf algebra perspective, the coproduct generates a shift 
along the braided line. 
To bring this out more clearly we adopt the notation of \cite{MajV}, 
and write 
\be
\eqalign{\theta&=1\otimes\theta\quad,\quad
\delta\theta=\epsilon=\theta\otimes1\quad,\cr}
\label{onetwo}
\ee
so that (\ref{nine}) leads to
\be
[\epsilon,\theta]_{q^{-1}}=0\quad.
\label{onethree}
\ee
Then, (\ref{seven}) can be written as
\be
\Delta\theta=\epsilon+\theta \quad,  
\label{onefour}
\ee
which, in standard supergroup notation, corresponds to the action of the left 
translation by $\epsilon$, 
$L_\epsilon\theta:\theta\mapsto\epsilon+\theta$.

The above expressions provide a basis upon which to construct 
a differential 
(and integral \cite{DMPA}) calculus on the braided line. 
We can introduce an algebraic left derivation operator $\d$, 
in analogy with the undeformed case, 
via the requirement
$[\epsilon \d,\theta]=\epsilon,$ 
which implies that 
\be
[\d,\theta]_q=1\quad,\quad
(\drr\theta=1)\quad.
\label{onesix}
\ee
At this stage $\d$ is still unspecified beyond (\ref{onesix}) but, as a 
{\it derivation}, it has dimensions inverse to those of $\theta$.
Regarding (\ref{onefour}) as the definition the 
left translation by $\epsilon$, we can go on to
consider right shifts $R_\eta:\theta\mapsto\theta+\eta$ 
of odd parameter $\eta$ where 
$[\theta,\eta]_{q^{-1}}=[\eta,\theta]_q=0$. 
Considerations similar to the above lead us to a right derivative operator 
$\dR$,
which satisfies 
\be
[\theta,\dR]_q=1\quad.
\label{oneseven}
\ee
The left and right derivative operators are in general related \cite{DMPA} by
\be
\dR=-q^{-(1+N)}\d\quad,
\label{oneeight}
\ee 
where $N$ is an operator satisfying,
\be
[N,\theta]=\theta\quad,\quad 
[N,\d]=-\d\quad,
\label{onenine}
\ee
and consequently $[N,\dR]=-\dR$. 
This implies that $ [\d,\dR]_{q^{-1}}=0.$
We will usually work with left derivatives. 
Let $f(\theta)$ be a function of $\theta$ defined by the positive power 
series expansion,
\be 
f(\theta)=\ss {C_m\theta^m\over[m]_q!}\quad,
\label{twoone}
\ee
where the $C_m$ are ordinary numbers. The derivative of $f(\theta)$ is 
generated by the graded bracket (\ref{onesix}) as follows,
\be
\eqalign{\drr \ff&=[\d,\ff]_z
=\ss C_m\left[\d ,\sp{\theta^m\over[m]_q!}\right]_z
=\ss C_m\left[\d ,\sp{\theta^m\over[m]_q!}\right]_{q^m}\cr
&=\ss C_m{\theta^{m-1}\over[m-1]_q!}\quad.\cr}
\label{twotwo}
\ee
This clearly reduces to the calculus on the undeformed line when $q=1$. 
An important point here is
that the derivative $\d$ is not a part of the algebra preserved by the 
coproduct, but rather the first order form of that coproduct. 
The full coproduct is related to a deformed
exponentiation of $\epsilon \d$ \cite{DMPA,AM,MajIV}.
From now on we refer to the differential
calculus defined by (\ref{onesix}), (\ref{oneseven}) and 
(\ref{oneeight}) as $q$-calculus.

\section{$q$-calculus in the $q\to-1$ limit and supersymmetry}
If we pass in a suitable way to the limit in which $q\to-1$, 
the $q$-calculus takes on a particularly interesting form, 
the full structure of which
has not to our knowledge been developed previously. In fact, we  take the 
limit  by setting 
$q=-1+iy$ and letting $y\to0$. 
Taking the limit along some other line would only change our results by a
phase factor, and this
could be cancelled out by using a different normalization for $t$ in 
(\ref{twoseven})
below. 
To take this limit we note that for generic $q$ we have the relationship
\be
\left[\d,{\theta^m\over[m]_q!}\right]_{q^m}={\theta^{m-1}\over[m-1]_q!}\quad,
\label{twofive}
\ee
where $m$ is a positive integer. In taking the $\li$ limit of this result 
we first encounter difficulties when $m=2$, because in this limit $[2]_q=0$. 
We can only retain (\ref{twofive}) 
in this limit if its LHS is kept finite and nonzero. 
This can be arranged by requiring that the  $\li$ limit of 
$\theta^2\over[2]_q!$  be finite and nonzero. 
Since $[2]_q=0$ this can only be the case if
$\theta^2\to0$ as $q\to-1$. 
To see that this is preserved under the left shift
$\theta\to\theta+\epsilon$, we note that  
when $\li$ (\ref{onethree}) becomes the familiar anticommutator
$\lbrace\epsilon,\theta\rbrace=0,$
so that 
$(\theta+\epsilon)^2=0,$
provided that $\theta^2$=$\epsilon^2$=0. 
We now note that under complex conjugation on the line
$q=-1+iy$,
we have $\overline{[2]_q!}=-[2]_q!$. 
From this it is clear that if we define 
\be
t:=\qm{i\theta^2\over[2]_q!}
\label{twoseven}
\ee
and, as assumed, $\theta$ is real, then $t$ will also be real. 
By using the following identities, valid for integer $r$,
\be
\eqalign{\qm\left({[2r]_q\over[2]_q}\right)=\qm
\left({1-q^{2r}\over1-q^2}\right)=r
\quad,}
\label{twoeight}
\ee
and 
\be
\eqalign{\qm\left({[2r+1]_q\over[1]_q}\right)&=\qm\left({1-q^{2r+1}\over1-q}
\right)=1\quad,}
\label{twonine}
\ee
we have, for $p=0,1$ (in which case $[p]_{-1}!=p!=1$) and integer $r$,
\be
\eqalign{\qm\left({\theta^{2r+p}\over[2r+p]_q!}\right)=
{\theta^p\over[p]_{-1}!}{(-it)^r\over r!}=
{\theta^p\over p!}{(-it)^r\over r!}\quad.}
\label{thirty}
\ee
The finite limit in (\ref{twoseven}) 
denoted by $t$ was introduced in order to handle 
difficulties encountered when considering the $\li$ limit of (\ref{twofive}) 
at $m=2$. 
Similar problems arise for all $m\geq2$,
and the importance of (\ref{thirty}) is that it tells us that these can also 
be handled in terms of $t$.
Eq. (\ref{thirty})
also tells us that any function $f(\theta)$ on the braided 
line at generic $q$ reduces to a function (`superfield') of the form 
$f(t,\theta)$ in the $q\to-1$
limit. 
To investigate further the properties of $t$, and to see how it fits into our
$q$-calculus, we now consider
\be
\left[\d,\left[\d,\sp{i\theta^2\over[2]_q!}\right]_{q^2}\right]_q=i\quad,
\label{threeone}
\ee
valid 
(cf. (\ref{twofive}))
for all $q\neq-1$. Taking the $\li$ limit we see that this reduces to
$[\d^2,t]=i,$
so by defining
\be
\dz=-i\d^2\quad,\quad
(\{\d,\d\}=2i\dz)\quad,
\label{threethree}
\ee
we have
\be
[\dz,t]=1\quad,\label{threefour}
\ee
which is just the defining relation of the algebra associated with ordinary 
calculus. 
From (\ref{threefour}) it can be seen that there are alternative 
ways of preserving (\ref{threeone}) 
in the $\li$ limit, corresponding to different
distributions of the ${1\over[2]_q!}$ factor between the definitions of $\dz$ 
and $t$. 
However apart from our choice, none of these 
permit introducing an operator reciprocal to $\d$, which 
is necessary if we are to 
define integration as in the generic $q$ case \cite{CZ,KM} so that it inverts the effect
 of differentiation. 
A discussion of $q$-integration when $q$ is an arbitrary root of unity will be given in 
\cite{DMPA}.
Using $\dz$ given by (\ref{threethree})
to induce differentiation with respect to $t$, the full $q$ calculus for $q=-1$ 
 obtained from (\ref{twotwo}) and (\ref{twoseven}) is given by,
\be
\eqalign{\drr\theta&=\lbrace \d,\theta\rbrace=1\quad,\quad
\drr t=[\d,t]=i\theta\quad,\cr
\dt t
&=[\dz,t]=1\quad,\quad \dt\theta=[\dz,\theta]=-i[\d,\{\d,\theta\}]=0\quad.\cr}
\label{threefive}
\ee
Since $\dt\theta=0$ and $\drr t\neq0$, we can only avoid a contradiction 
by interpreting $\dt$ as a partial derivative, and $\drr$ as a total 
derivative, a result which we implicitly took into account when
choosing our notation. 
We can also define partial differentiation with respect to $\theta$. We do
this as follows
\be
\eqalign{\pt\theta &:=\{\partial_\theta,\theta\}=1\quad,\quad
\pt t:=[\partial_\theta,t]=0\quad.
\cr}
\label{threesix}
\ee
Using this definition we are able to perform a chain rule expansion of the 
total derivative, so that
\be
\eqalign{\drr={d\theta\over d\theta}\pt+{d t\over d\theta}\dt=\pt+i\theta\dt
\quad.\cr}
\label{threeseven}
\ee
By substituting (\ref{threeseven}) into the definition (\ref{threethree}) 
we obtain an additional but expected condition,
\be
{\partial^2\over\partial\theta^2}=0\quad.
\label{threeeight}
\ee
This can all be put into the algebraic form $\d=\partial_\theta+i\theta\dz,$ 
where
$\lbrace\partial_\theta,\theta\rbrace=1$
and
$\partial_\theta^2=0.$
Now, let us introduce the operator $q^{-N}$ where $N$ is defined as in 
(\ref{onenine}). 
When $q=-1$ we can write 
$q^{-N}=(-1)^{-N}=1-2\theta\d=1-2\theta\partial_\theta$ \cite{DMPA}. 
Then if we define 
\be
\eqalign{Q=\d\quad,\quad
D=\dR=(-1)^{-N} \d\quad,\cr}
\label{fourfour}
\ee
where we have made use of (\ref{oneeight}), we have
\be
\eqalign{Q=\partial_\theta+i\theta\dz\quad,\quad
D=\partial_\theta-i\theta\dz\quad,\cr}
\label{fourfive}
\ee
as well as 
\be
\lbrace Q,D\rbrace=0\quad,\quad
Q^2=i\dz\quad,\quad 
(\{Q,Q\}=2i\partial_t)\quad,\quad
D^2=-i\dz\quad,\quad
(\{D,D\}=-2i\partial_t)
\quad.
\label{foursix}
\ee 
From (\ref{foursix}) it is clear that $Q$ is just the  supercharge encountered 
in one-dimensional supersymmetry, and $D$ the corresponding covariant 
derivative; the picture, however, remains the same when spacetime is not 
one-dimensional. 
Noting that $\partial_\theta$ and
$\theta$ satisfy the algebra associated with ordinary Grassmann variables, 
and identifying $t$ as the time variable, we see that the usual  
superspace realizations of $Q$ and $D$ are given by (\ref{fourfive}). 
However our derivation is new, and from 
it we can see that the properties associated with these operators stem 
from underlying {\it total} left and right derivatives.
Using (\ref{fourfour}) we can now identify $Q$ and $D$ as,
respectively, the generators of left and right shifts along the braided 
line at $q=-1$
( In the (super)group language they may be identified with 
the right- $[Q]$ and left-invariant $[D]$  generators \cite{AldAz}).

To further investigate this point of view we examine the Hopf structure on the 
braided line in the $\li$ 
limit. When $q=-1$, (\ref{nine}) reduces to
\be
\eqalign{(1\otimes\theta)(\theta\otimes1)&=-\theta\otimes\theta\quad,\quad
(\theta\otimes1)(1\otimes\theta)=\theta\otimes\theta\quad;\cr}
\label{fourseven}
\ee
so that from (\ref{seven}), 
\be
\eqalign{\Delta\theta^2=\theta^2\otimes1+1\otimes\theta^2+(1\otimes\theta)
(\theta\otimes1)+(\theta\otimes1)(1\otimes\theta)
=0\quad,\cr}
\label{fournine}
\ee
as required by the homomorphism property of the coproduct.
The counit and antipode take the following form
\be
\eqalign{\varepsilon(\theta)&=0\quad,\quad
S(\theta)=-\theta\quad.
\cr}
\label{fifty}
\ee
Thus far nothing in this account of the Hopf structure is new, 
see for example \cite{MajIII} for a
discussion of super and anyonic quantum groups. 
The important point is that there is an additional,
previously undiscussed structure, which cannot be seen until the variable $t$ 
as defined by (\ref{twoseven}) is introduced. 
From (\ref{twoseven}) and (\ref{seven})-(\ref{nine}) 
we compute the $q\to-1$ limit of 
$\Delta [{i\theta^2\over[2]_q!}]$
and see using also (\ref{oneone})
that $t$ has the following braided Hopf 
structure,
\be
\Delta t=t\otimes1+1\otimes t+i\theta\otimes\theta\quad,\quad
\varepsilon(t)=0\quad,\quad
S(t)=-t\quad.
\label{fiveone}
\ee
This is an interesting result, because it means that although $t$ and $\dz$ 
satisfy the algebra associated with ordinary (undeformed) calculus, $t$ has 
non standard braided Hopf structure. 
From the definition (\ref{twoseven}) it follows that 
$\theta$ and $t$ commute. 
Considered along with the chain rule 
expansion of the $q$-calculus
derivative (\ref{threeseven}), it is clear that in the $\li$ limit the 
$q$-calculus algebra
can be written in terms of independent $t$ and $\theta$ parts.
However, from (\ref{fiveone}), we see that $\theta$ appears in the coproduct 
of $t$, and so the when we consider the full braided Hopf structure, no such 
separation can be performed. 
The fact that we cannot regard this braided Hopf algebra
as a product entity is fundamental to this view of supersymmetry. 
To see this we rewrite the coproducts of $\theta$ and $t$ using the notation 
of (\ref{onetwo})-(\ref{onefour}),
\be
\eqalign{\theta\to\Delta\theta=\epsilon+\theta\quad,\quad
t\to\Delta t=t+t_{\epsilon}+i\epsilon\theta\quad,\cr}
\label{fivetwo} 
\ee
where 
\be
t_{\epsilon}=\qm{i\epsilon^2\over[2]_q!}
\label{fivethree}
\ee
is the even variable associated with $\epsilon$ (cf. (\ref{twoseven})), 
and is independent of $t$.
Comparing this with (\ref{onea}) and (\ref{oneb}) we see that 
(\ref{fivetwo}) is just the usual form 
of a finite supersymmetry transformation; only now the transformation 
of $t$ follows from
that of $\theta$ via the relationship (\ref{twoseven}).

To conclude, let us summarize our results and
 outline the new point of view which they provide. 
For generic $q$ the braided line described in section 3 is well defined. 
In the non-trivial $\li$ limit employed here 
the nilpotency of the odd operator 
$\theta$ prevents it from providing a complete description of the braided line. 
A convenient way in which we can obtain
such a description is to introduce an additional variable $t$,
defined as in (\ref{twoseven}). 
From (\ref{thirty}) this is seen to carry structure which, 
for generic $q$, is related
to $\theta^2$ and higher powers of $\theta$. So in
the $q\to-1$ limit used here the braided line is made up
of the two variables $\theta$ and $t$, which span the one-dimensional 
superspace. 
Furthermore, under a shift along this braided line $\theta$ and $t$ transform 
exactly as in supersymmetry (\ref{fivetwo}). 
Thus we are able to identify superspace with the braided line 
in the limit in question here, 
and supersymmetry as translational invariance along this line. 
This means that
superspace cannot be regarded as the tensor product of independent 
$t$ and $\theta$ parts, but that it is
instead a single braided geometric entity.
From the point of view of this braided geometric interpretation,
supersymmetry is not a symmetry between independent odd and even sectors, 
but translational invariance within a single space.

Our work also provides a braided interpretation
of the ideas developed in \cite{AldAz}, where rigid superspace is viewed as 
the central extension of the odd supertranslation group by the group of 
ordinary (even) translations. 
Note also that the definitions (\ref{twoseven}), (\ref{threethree}), 
and the identification (\ref{fivethree}),
make exact the notion of superspace as a non-trivial square root 
of ordinary space.
\vskip30pt\noindent
\begin{Large}
{\bf Acknowledgements}
\end{Large}
\vskip10pt

This paper describes research supported in part by E.P.S.R.C and P.P.A.R.C. 
One of the authors (J.A.) wishes to thank the CICYT (Spain)
for a research grant. J.C.P.B. wishes to acknowledge an FPI grant
from the Spanish Ministry of Education and Science and the CSIC.

\end{document}